\documentclass[twocolumn,showpacs,amsmath,amssymb,prb]{revtex4-1}
\usepackage{graphicx}
\usepackage{bm}
\usepackage{dcolumn}

\begin{document}
\title{High temperature non-collinear magnetism in a classical bilinear-biquadratic Heisenberg model}

\author{Kanika Pasrija and Sanjeev Kumar}
\affiliation{Indian Institute of Science Education and Research(IISER) Mohali, Knowledge City, Sector 81, 
Mohali 140 306, India}

\date{\today}

\begin{abstract}
Motivated by the magnetically-driven high-temperature ferroelectric behavior of CuO and the subsequent theoretical efforts to understand this intriguing phenomenon, we study a spin model on a two-dimensional square lattice 
which possesses some of the key features of the models proposed for CuO. The model consists of Heisenberg couplings between nearest and next-nearest neighbor
spins, and biquadratic couplings between nearest neighbors. We use a combination of variational calculations and classical Monte Carlo simulations to study this
model at zero and finite temperatures. 
We show that even an arbitrarily weak biquadratic coupling plays a crucial role in selecting the magnetic ground state. More importantly, 
a non-collinear magnetic state, characterized by a finite spin current, is stable at finite temperatures.
The interesting aspect is that the present model neither includes an 
inversion-symmetry-breaking term nor the effects of lattice distortions in the Hamiltonian. 
We conclude that non-collinear magnetism at high temperatures, as observed in CuO, can be explained via 
pure spin Hamiltonians. We find that the spiral phase is inhomogeneous, and is stabilized by entropic effects.
Our study demonstrates
that higher order interaction terms are of crucial importance if the stronger interactions together with the lattice geometry contemplate to generate a near degeneracy of magnetic states.
The conclusions presented in this work are of particular relevance to the non-collinear magnetism and ferroelectricity observed at high temperatures in cupric oxide. 
\end{abstract}

\pacs{75.10.Hk, 75.40.Mg, 75.85.+t, 75.47.Lx}
\maketitle

\section{Introduction}

Materials exhibiting more than one ferroic ordering phenomena in a single phase are defined as multiferroics \cite{Schmid-94}. Over the last few years, magneto-electric
multiferroics have received a special attention from solid-state and materials researchers for the exciting possibility of controlling magnetization by external electric field and electric polarization by external 
magnetic field \cite{magnetoelectrics}.
It was soon realized that there is a scarcity of materials that simultaneously exhibit ferroelectricity and ferromagnetism \cite{BMO, Rabe}. In fact, it appeared that a ferroelectric order rarely coexists with
any long-range magnetic order \cite{Hill, Hill02}. This was attributed to the fact that ferroelectric materials typically have a $d^0$ electronic configuration, whereas it is the partial filling of $d$-levels that is
responsible for magnetism in a large number of compounds. Following this observation researchers started to search for materials that are exceptions to this {\it rule}.
Over the course of last decade, the original definition of magneto-electric multiferroic order has been extended to include 
materials which exhibit ferroelectric order together with any long-range magnetic order \cite{Review}. Within this new definition of multiferroics, a large class of 
new multiferroic compounds have been discovered, which range from transition-metal oxides to organic crystals \cite{Perks-12, Ishiwata, Chapon, Kimura-06, Seki, Tokura-08, Loidl-12}.
On the theoretical side, many new mechanisms were proposed that allow the ferroelectric order to coexist with a magnetic order \cite{JvdB-08, Katsura, Mostovoy, Solovyev}.

One of the motivations for such immense interest in this research field is the possibility of designing powerful memory devices using multiferroic materials, where an electrical writing and a non-destructive
magnetic reading is allowed \cite{Scott, Ramesh, Cheong}. Therefore, switching magnetization (polarization) by an external electric (magnetic) field are the most desirable characteristics for a useful multiferroic material.
The necessary condition for this special feature to be present in multiferroics is the existence of
a strong coupling between the ferroelectric and magnetic order parameters. Therefore one guiding principle has been to search for systems
where the magnetic and ferroelectric orders are not independent of each other. 
Strong magneto-electric coupling was discovered in some systems with a spiral type long-range magnetic order, where
the ferroelectricity is induced by the spiral spin states \cite{spiral-MF}. Typically, such spiral spin states emerge either from competing interactions or from the Dzyaloshinskii-Moriya (DM) interactions, and therefore
these states exist only at low temperatures. Indeed, such spiral magnetic phases have been observed below $\sim 40K$ in some manganites \cite{TbMnO}. The ferroelectrics where the magnetic order is responsible
for the ferroelectricity have come to be known as improper ferroelectrics \cite{improperFE}.
Incidentally, there have been very few examples of systems where a magnetic order is induced by a ferroelectric order \cite{weakFM}.

While the low-temperature multiferroics help us in understanding
various mechanisms, they are far from being useful in any realistic applications. Therefore, the second challenge in this field is to find a multiferroic material with high transition temperatures.
Although, the most well known multiferroic, BiFeO$_3$, has large magnetic and ferroelectric ordering temperatures, the coupling between the two order parameters is rather weak \cite{BFO}.
This is due to the fact that the magnetism in BiFeO$_3$ comes from the Fe, whereas ferroelectricity arises due to off-centering of the Bi ions. The temperature scales for magnetic and ferroelectric ordering
are very different, which is also an indication of the independent origins of the two ordering phenomena in this material.
Interestingly, it is an oxide of Cu that emerges as a promising candidate with high transition temperatures and strong magnetoelectric coupling.

Oxides of Cu have a very special place in the history of condensed matter physics thanks to the famous high-temperature superconductors, the cuprates \cite{cuprates}. 
Another oxide of Cu, cupric oxide (CuO), has been of research interest due to its intriguing magnetic properties \cite{CuO-old, Yang, Choi, orbital-current}. 
Synthesis of CuO in tetragonal phase has caused much excitement in the scientific community for its possibly large value of Neel temperature $T_N$ \cite{tetragonal-CuO}.
Recent discovery of a ferroelectric behavior in CuO at $\sim 230$K has revived research interest in 
this material from the point of view of multiferroics \cite{CuO-Kimura}. 
In fact, it is the 'high-Tc' material as far as the multiferroics with large magneto-electric coupling are concerned \cite{high-Tc}. 
The only disadvantage appears to be the narrow range of existence of the ferroelectric order, which coincides with a non-collinear magnetic order observed between $213$K and $230$K.
However, recent theoretical studies have predicted that the range of stability as well as the Neel temperature of the non-collinear magnetic order can be tremendously increased by external pressure, leading to 
a possible room-temperature multiferroic behavior in CuO \cite{CuO-pressure}.

The discovery of this high-Tc multiferroic material posed various questions regarding the microscopic understanding of the phenomena. 
A number of theoretical investigations have been reported since the discovery of high-temperature ferroelectricity in CuO \cite{CuO, He, XR, Toledano}. It has been established that the magnetic interactions in CuO are 
rather complicated, 
and lead to competing ground states. There is still no consensus on the experimental values of the various exchange parameters \cite{CuO-old,Yang,Choi}. Similarly, the outcome of the theoretical calculations of the exchange 
parameters depend on the details of the method used \cite{CuO, XR, Cao}. Nevertheless, most theoretical models point to a near degeneracy of 
magnetic states in this system.
Giovannetti {\it et al.} proposed that the competition
between a small uniaxial anisotropy and weak DM interactions leads to the observed experimental behavior of the ferroelectric polarization. The model was further improved by Jin {\it et al.} by explicitly 
including the lattice parameters in the microscopic Hamiltonian. Both these models point towards a crucial role played by entropic effects in stabilizing the non-collinear magnetic phase at high temperatures. 

The high temperature crystal structure of CuO is centrosymmetric and the inversion symmetry breaking occurs at the magnetic ordering temperature. In fact, the non-collinear magnetic order explicitly breaks the
inversion symmetry and the symmetry breaking in the crystal structure can be seen as a secondary effect via the inverse DM mechanism \cite{iDM}. Therefore, it is important to ask if the non-collinear magnetic order observed in
CuO can be explained via pure spin Hamiltonians arising out of electron-electron interactions, without including the effects of lattice distortions. The ferroelectric distortions can then be understood in 
terms of the inverse DM effect.

In this work we explore some features of the magnetic ordering phenomena in CuO without taking into account the detailed crystal
structure of the material. In order to keep
the scope of this work general we use a simple two dimensional (2D) square lattice with nearest and next-nearest neighbor magnetic exchange couplings. 
The model does not include an explicit symmetry breaking term such as a DM interaction, however it includes 
higher order spin-spin interaction terms. Such terms arise in the fourth order in strong coupling expansion of the Hubbard Hamiltonian. Although the strength of the higher order terms is much weaker than 
the Heisenberg exchange terms, 
the former become important due to near degeneracy of states in the pure Heisenberg model. 
We present extensive Monte Carlo simulation results on a 2D bilinear-biquadratic Heisenberg model.
We find that the biquadratic coupling drives a first order phase transition between a collinear up-up-down-down type antiferromagnet (denoted as EAF) and a spiral antiferromagnet at low temperatures. 
At finite temperatures the non-collinear phase is found to be more stable compared to the collinear phase. This 
leads to a sequence of phase transitions from a high-temperature paramagnetic state to a non-collinear state, and finally to a collinear EAF phase at low temperatures. 
This is precisely the behavior reported in the
experimental studies of CuO. we find that a pure electronic Hamiltonian is sufficient to describe the magnetism in CuO provided higher-order interaction terms are taken into account.
While our model study is of general importance for magnetic materials and models, the conclusions are of particular 
relevance to the magnetic and ferroelectric ordering phenomena in CuO.

The rest of the paper is organized as follows. In section II we describe possible extensions of the standard Heisenberg model for spin-$1/2$ and spin-$1$ magnetic moments.
We discuss the relevance of these models and their relation with each other in the classical limit. Section II ends with a description of the
variational, and Monte Carlo (MC) simulation methods used in this work.
Section III begins with a discussion of the ground state phase diagrams of the two models using variational calculations. Next, a comparison between variational calculations and MC simulations at low temperatures 
is presented. Finally, the finite-temperature behavior of the bilinear-biquadratic model is discussed.
The main focus is on the
presence of a 
non-collinear magnetic phase with finite spin current at high temperatures. 
Conclusions are presented in Section IV.

\section{Model and Method}

We begin with a Heisenberg Hamiltonian on a square lattice given by,

\begin{eqnarray}
H_0 &=& J_1 \sum_{ \langle ij \rangle} {\bf S}_i\cdot{\bf S}_j + J_{2a} \sum_{ \langle \langle ij \rangle \rangle_a} {\bf S}_i\cdot{\bf S}_j
+ J_{2b} \sum_{ \langle \langle ij \rangle \rangle_b} {\bf S}_i\cdot{\bf S}_j. \nonumber  \\
\end{eqnarray}

\begin{figure}
\includegraphics[width=.95\columnwidth,angle=0, clip = 'True' ]{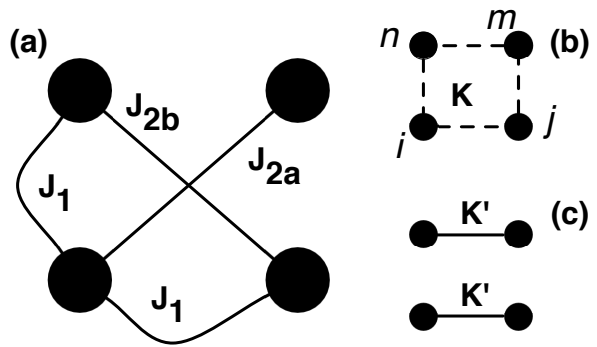}
\caption{Schematic representation of various interactions that constitute the model Hamiltonians $H_1$ and $H_2$ discussed in the text. 
(a) The nearest-neighbor, and the next-nearest-neighbor Heisenberg exchange couplings on a square lattice,
(b) the four-spin ring-exchange coupling in Hamiltonian Eq. (2), and (c) the biquadratic coupling in Hamiltonian Eq. (3).
}
\label{fig1}
\end{figure}

\noindent
Here, $J_1$ denotes the nearest neighbor (nn) Heisenberg exchange coupling and $J_{2a}$, $J_{2b}$ are the next nearest neighbor (nnn) couplings as shown in Fig. \ref{fig1}(a). The
single and double angular brackets denote the sum over nn and nnn sites, respectively. The subscripts $a$ and $b$ on the summation indices specify the two inequivalent nnn directions.
For most spin systems with square lattice geometry,
one is typically interested in the parameter regime given by, $|J_{2a}| \sim |J_{2b}| \leq |J_1|$. However, another 
interesting limit of this model is realized when $J_{2a}$ and $J_{2b}$ have opposite signs and are much larger
in magnitude compared to $|J_1|$. The corresponding model for Ising spins was 
recently analyzed by A. Kalz and G. Chitov, and the existence of an unusual topological floating phase was reported \cite{Ising-J1J2}.
It is easy to see that in this limit, the magnetic system gets divided into two sub-lattices which interpenetrate each other. $J_{2a}$ and $J_{2b}$ ensure that each sub-lattice
has a well defined order at low temperatures, but the nn Heisenberg coupling $J_1$ is not sufficient to generate a long-range magnetic order. Therefore, in this parameter regime higher order
spin-spin interaction terms become relevant. Interestingly, the magnetic model for CuO corresponds to a similar sub-lattice order in three dimensions (3D) where each sub-lattice has a well defined order but the 
magnetic ground state is degenerate and additional weaker couplings, such as the magneto-crystalline anisotropy and the Dzyaloshinskii-Moriya (DM) interaction, are considered important. 
However,  the higher order spin-spin
interactions should be given equal importance in this regime. Proceeding with this viewpoint,
our first extension of the Heisenberg Hamiltonian
$H_0$ is achieved by including a four-spin ring-exchange interaction, leading to the Hamiltonian,

\begin{eqnarray}
H_1 &=& H_0 + K \sum_{[ijmn]} \left[ ({\bf S}_i\cdot{\bf S}_j) ({\bf S}_m\cdot{\bf S}_n) + \right. \nonumber \\ 
& & \left.   ({\bf S}_i\cdot{\bf S}_n) ({\bf S}_j\cdot{\bf S}_m) -({\bf S}_i\cdot{\bf S}_m) ({\bf S}_j\cdot{\bf S}_n) \right ].
\end{eqnarray}

\noindent
In the above, $K$ denotes the strength of the ring-exchange coupling involving four sites (see Fig. \ref{fig1}(b)). 
Starting with a one-band Hubbard model at half-filling, the 2nd order perturbation theory in 
hopping leads to the Heisenberg exchange \cite{Kugel-Khomskii}. If we go
beyond the second order, the next contribution is from the fourth order term leading to the ring-exchange coupling \cite{ring_exchange}. 
Therefore, $H_1$ is the microscopic Hamiltonian for spin-$1/2$ moments on a square lattice. 
If the magnetic moments are spin-$1$, then the next order term is a biquadratic one involving two sites \cite{Mila-biquad}. We define our 2nd Hamiltonian, $H_2$, by including a biquadratic term to $H_0$. The
Hamiltonian is given by,

\begin{eqnarray}
H_2 &=& H_0 + K' \sum_{ \langle ij \rangle_x} ({\bf S}_i \cdot {\bf S}_j)^2 .
\end{eqnarray}

\noindent
Here, $K'$ is the coupling strength of the biquadratic interaction, which we have taken
to be present only on the nn bond along the x-direction (see Fig. \ref{fig1}(c)).

$H_1$ ($H_2$) is the spin Hamiltonian containing interaction terms up to 4th order for spin-$1/2$ (spin-$1$) magnetic moments. However, in the classical limit these two models
have many similarities. This is particularly easy to recognize when $J_{2a}$ and $J_{2b}$ are stronger couplings compared to $J_1$. The term inside the square brackets in Eq. (2) can
be rewritten as $[({\bf S}_i\cdot{\bf S}_j) ({\bf S}_m\cdot{\bf S}_n) - ({\bf S}_i \times {\bf S}_j) \cdot ({\bf S}_m \times {\bf S}_n) ]$. Now given that the nnn couplings are much stronger, at low temperatures
${\bf S}_i$ (${\bf S}_j$) is either parallel or antiparallel to ${\bf S}_m$ (${\bf S}_n$), depending on the sign of $J_{2a}$ ($J_{2b}$). 
Making the substitutions ${\bf S}_m \rightarrow {\bf S}_i$ and ${\bf S}_n \rightarrow {\bf S}_j$, we can
rewrite $H_1$ as,

\begin{eqnarray}
H_1 &=& H_0 + K \sum_{\langle ij \rangle} \left[ ({\bf S}_i\cdot{\bf S}_j)^2  - ({\bf S}_i \times {\bf S}_j)^2 \right ].
\end{eqnarray}

We finally note that $({\bf S}_i \times {\bf S}_j)^2 = |S|^4 - ({\bf S}_i\cdot{\bf S}_j)^2$, and therefore the two models $H_1$ and $H_2$ are identical up to an additive constant and the sign of the coupling constant. 
The equivalence of
these two Hamiltonians is valid only in the classical limit, and at temperatures scales much smaller than the nnn coupling strengths.

\begin{figure}
\includegraphics[width=.95\columnwidth,angle=0, clip = 'True' ]{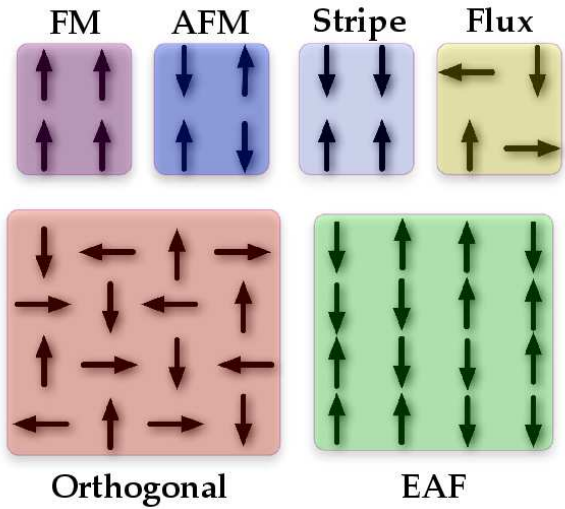}
\caption{(Color online) Schematic view of various magnetically ordered states considered in the variational calculations. A general spiral phase is not shown here. The ferromagnet, antiferromagnet, 
stripe and orthogonal 
phases are limiting cases of spiral states with $q_x = q_y = 0$ (FM), $q_x = q_y = \pi$ (AFM), $q_x = 0, q_y = \pi$ (stripe) and $q_x = q_y = \pi/2$ (orthogonal).
The Flux phase and the up-up-down-down type antiferromaget, EAF, do not belong to the set of spiral states.
}
\label{fig2}
\end{figure}

The bilinear-biquadratic Heisenberg model has been well studied as a quantum spin-$1$ Hamiltonian on various lattices \cite{Biquad1,Biquad2,Biquad3}. 
The model has also been studied recently in the classical limit \cite{Mila}, where the
focus was on the existence of non-coplanar phases, and the ground-state degeneracies on a triangular lattice.
The model has also been studied on pyrochlore lattice in the context of Cr spinels and their magnetic properties \cite{Shannon}. In addition, there have been
numerous studies trying to understand the properties of quantum spin chains using a variety of theoretical methods \cite{spin-chain}.
In this paper, we study the bilinear-biquadratic Heisenberg Hamiltonian as a candidate for non-collinear magnetism at high-temperatures in systems where the pure Heisenberg term leads to
degenerate ground states. 

In the next section we compare the classical ground state phase diagrams of the two Hamiltonians, $H_1$ and $H_2$. The parameters of the models are, $J_1$, $J_{2a}$, $J_{2b}$ and $K(K')$. Since we are 
interested in the regime where nnn couplings are stronger than nn ones, we set $J_{2a} = 1$ as our energy scale, and all other
couplings and energies are therefore expressed in terms of $J_{2a}$. 
In order to compare the ground state phase diagrams of the two models described by $H_1$ and $H_2$, we use a variational scheme. We consider planar spiral spin states with spiral wave-vector ${\bf q} = (q_x, q_y)$.
The orientation of the magnetic moment at lattice site ${\bf r}_i$ is given by, 
${\bf S}_i = |S|(\cos ({\bf q} \cdot {\bf r}_i), \sin ({\bf q} \cdot {\bf r}_i) , 0)$. Such spiral states contain the conventional magnetically ordered states, 
such as the ferromagnet, staggered antiferromaget, stripe-type antiferromaget and orthogonal state as limiting cases (see Fig. \ref{fig2}). We minimize the total energy of a general spiral state for different values of model parameters. 
In addition, we include two more states in our minimization scheme. These are the so called flux state and the up-up-down-down type antiferromaget (EAF) state (see Fig. \ref{fig2}). We make use of mathematica for 
performing the variational minimization described above.

Classical Monte Carlo simulations are employed for the finite-temperature study of the bilinear-biquadratic Hamiltonian \cite{Newman_book}.
The standard Markov Chain Monte Carlo method is used with the Metropolis algorithm for configuration updates \cite{Metropolis}. Single-spin update moves are performed by randomly selecting a pair of polar and azimuthal angles for a given spin.
The move is accepted with the Boltzmann probability $e^{-\Delta E / (k_B T)}$, where $\Delta E$ is the energy difference between the new and the old configurations, $T$ is the temperature and $k_B$ is the Boltzmann constant.
Following the standard practice in such simulations, we set $k_B = 1$ so that the temperature scales become equivalent to the energy scales.
Each spin is updated $N_{eq} \sim 10^6$ times for the purpose of thermalization of the system. Thereafter, we begin to compute physical quantities which are averaged over 
$N_{av} \sim 10^6$ further update steps. Most of the results are presented on a square lattice with $120^2$ spins. However, the stability of results is checked on lattice sizes varying 
from $80^2$ to $200^2$ for selected parameter values.

\section{Results and Discussions}

\subsection{Ground state Phase Diagrams}

We begin by presenting the classical ground state phase diagrams for the Hamiltonian Eq. (2). Due to the presence of a four-spin interaction term, the standard method of Luttinger and Tisza for finding the 
ground states of a classical spin
Hamiltonian does not work \cite{FT_method}. Therefore, we rely on variational calculation for determining the ground state magnetic phases of the model. 
We consider the general spiral states, the up-up-down-down type antiferromagnetic 
(EAF) state and the flux state, as described in the previous section. We also allowed for spiral states with a finite ferromagnetic moment, but such phases were not stable for any combination of
the parameter values explored. A schematic view of the different ordering patterns considered in our variational calculations is shown in Fig. \ref{fig2}. 
The phase diagrams are obtained by minimizing the energy of the system over these states.
As mentioned earlier, we set $J_{2a} = 1$ as the reference energy scale. Therefore
the free parameters to be explored are $J_1$, $J_{2b}$ and $K$. 

The energy per spin for various ordered magnetic states for the Hamiltonian Eq. (2) is given by the following equations: 

\begin{eqnarray}
E_{EAF} & = & -J_{2a} + J_{2b} - K \\
E_{Spiral} & = & J_1 (\cos q_x + \cos q_y) + J_{2a} \cos (q_x + q_y) \nonumber \\  
     & & + J_{2b} \cos (q_x - q_y) + K \left [ \cos^2 q_x + \cos^2 q_y \right . \nonumber  \\
     & & \left . - \cos (q_x + q_y) \cos (q_x - q_y) \right ]  \\
E_{Flux} & = & -J_{2a} - J_{2b} - K 
\end{eqnarray}

\begin{figure}
\includegraphics[width=.99\columnwidth,angle=0, clip = 'True' ]{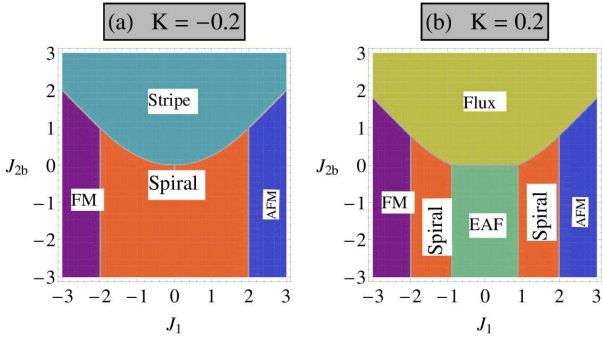}
\caption{(Color online) $J_1$-$J_{2b}$ ground state phase diagram of the Hamiltonian $H_1$ for the ring-exchange coupling values, (a) $K=-0.2$ and (b) $K= 0.2$. For $K=-0.2$ and $J_{2b} < 0$, the FM phase 
continuously evolves to a staggered antiferromagnetic
phase via spin spiral magnetic states with the wave-vector following the condition $q_x = q_y$.
}
\label{fig3}
\end{figure}

\begin{figure}
\includegraphics[width=.99\columnwidth,angle=0, clip = 'True' ]{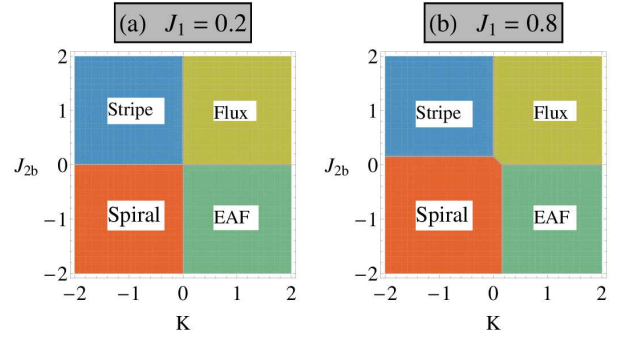}
\caption{(Color online) $K$-$J_{2b}$ ground state phase diagram of the Hamiltonian $H_1$ for two values of the nn Heisenberg exchange coupling, (a) $J_1=0.2$ and (b) $J_1= 0.8$.
}
\label{fig4}
\end{figure}

In Fig. \ref{fig3}, we show the $J_1$-$J_{2b}$ phase diagrams for two values of the ring-exchange coupling strength $K$. The different phases are indicated with acronyms in the figure. 
For negative values of $K$, all ground state phases are of spiral type. The region denoted as Spiral in the phase diagram consists of spiral phases with $q_x = q_y$. 
The FM state is continuously connected to AFM via the spiral states.
For positive values of $K$, the flux state and the EAF state occupy a considerable region of the parameter space. The boundaries between the spiral phase and the EAF phase are first-order in nature since the order
parameter of one phase is not continuously connected to that of the other.
The two most relevant phases
are the EAF and the orthogonal (spiral with $q_x = q_y = \pi/2$) phase. These are the 2D analogs of the
collinear AF$_1$ and non-collinear AF$_2$ states observed in CuO \cite{CuO-kimura}.
In Fig. \ref{fig4}, we show the $K$-$J_{2b}$ phase diagrams at two values of the nearest neighbor coupling $J_1$. The four stable phases are almost symmetrically placed in four quadrants for a small value of $J_1$.
Increasing $J_1$ leads to an enhanced stability of the spiral phase. Note that the spiral phase is of importance in the context of improper multiferroics, since it supports a finite spin current.

Having discussed the phase diagrams for the Hamiltonian with ring-exchange coupling, we now turn to the bilinear-biquadratic Hamiltonian specified by Eq. (3). 
The expressions for energy per spin of different ordered states is given by the following equations:

\begin{eqnarray}
 E_{EAF} & = & -J_{2a} + J_{2b} + K' \\
 E_{Spiral} & = & J_1 (\cos q_x + \cos q_y) + J_{2a} \cos (q_x + q_y)  \nonumber \\
     & & + J_{2b} \cos (q_x - q_y) + K' \cos^2 q_x   \\
 E_{Flux} & = & -J_{2a} - J_{2b} 
\end{eqnarray}

\begin{figure}
\includegraphics[width=.99\columnwidth,angle=0, clip = 'True' ]{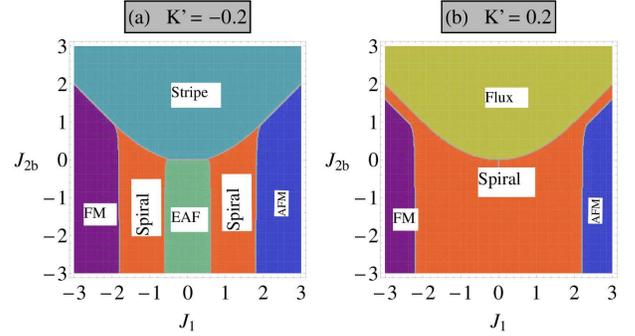}
\caption{(Color online) $J_1$-$J_{2b}$ ground state phase diagram of the Hamiltonian $H_2$ for two values of the biquadratic coupling strength, (a) $K'=-0.2$ and (b) $K'= 0.2$.
Note the similarity with the phase diagram shown in Fig. \ref{fig3}.
}
\label{fig5}
\end{figure}

\begin{figure}
\includegraphics[width=.99\columnwidth,angle=0, clip = 'True' ]{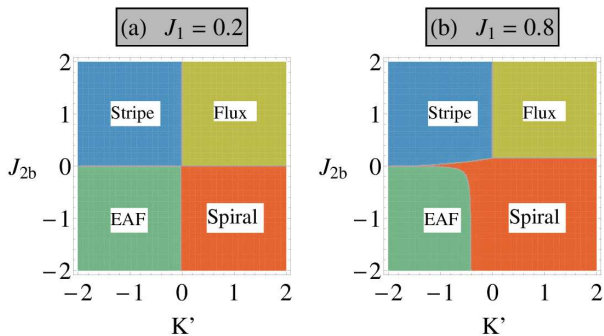}
\caption{(Color online) $K'$-$J_{2b}$ ground state phase diagram of the bilinear-biquadratic Hamiltonian $H_2$ for, (a) $J_1=0.2$ and (b) $J_1= 0.8$.
Similarity with the phase diagrams shown in Fig. \ref{fig4} is apparent.
}
\label{fig6}
\end{figure}

Once again, we use a variational approach to obtain the phase diagrams. The model parameters in this case are $J_1$, $J_{2b}$ and $K'$.
In Fig. \ref{fig5}, we show the $J_1$-$J_{2b}$ phase diagrams at two values of the biquadratic coupling strength $K'$. 
The different phases are again indicated with acronyms in the figure. 
There is a remarkable similarity between these phase diagrams and those shown in Fig. \ref{fig3}. Both sets of phase diagrams contain identical phases with almost
identically located phase boundaries, except for a relative sign between the couplings $K$ and $K'$ for negative $J_{2b}$ as discussed in Section II.
Once again, the relevant phases having collinear and 
non-collinear magnetic orders are part of the ground state phase diagram. 
\noindent
To complete the comparison we present the $K'$-$J_{2b}$ phase diagram for two values of $J_1$ in Fig. \ref{fig6}. Once again the phase diagrams contain the same four phases as
shown in Fig. \ref{fig4}. In addition, increasing $J_1$ leads to
an expansion of the stability regime for the spiral states.

The variational results highlight the rich behavior of the models $H_1$ and $H_2$ in terms of the presence of a variety of phases and phase boundaries.
The results further suggest that while the form of higher-order spin-spin interactions arising from a strong-coupling expansion is different for spin-$1/2$ and spin-$1$ magnetic moments,
their classical versions lead to phase diagrams with strong similarities. Given that the bilinear-biquadratic Hamiltonian has a simple form with each term containing pairwise interactions between spins, we select $H_2$ for finite-temperature studies.
Since the magnetic moments in CuO are spin-$1/2$, a microscopic biquadratic coupling does not exit. However, under the circumstances discussed in section II, the results of a bilinear-biquadratic
model can be relevant to spin-$1/2$ systems as well, specially in the classical limit.
We now proceed to study $H_2$ in more detail. Our main aim is to understand the finite-temperature
behavior of a Heisenberg model whose ground state degeneracy is lifted by a higher-order spin-spin coupling term.

\subsection{Comparison between variational and Monte Carlo results}

The variational phase diagrams discussed in the previous subsection show that the model contains a variety of phases and phase transitions in the ground state. In order to ensure that we have not missed any
crucial magnetic state in our variational set up, we verify the stability of some of these phases using Monte Carlo simulations. The details of the method are given in section II. 
We present the Monte Carlo results on the bilinear-biquadratic Hamiltonian given by Eq. (3).
Various quantities calculated in the Monte Carlo simulations are defined in the following. 
The total energy per spin, $E = \frac{1}{N}\langle H \rangle$, is computed by taking the thermal average of the Hamiltonian over the Monte-Carlo generated finite-temperature configurations. Here onwards, 
the angular brackets
denote thermal average, which for any quantity $X$ is defined as,

\begin{equation}
\langle X \rangle = \frac{1}{N_{av}} \sum_{\alpha=1}^{N_{av}} X_{\alpha},
\end{equation}

\noindent
where $\alpha$ denotes a single configuration with
$X_{\alpha}$ being the value of the quantity $X$ for that configuration and $N_{av}$ is the number of MC steps over which the averaging is performed.
In order to characterize the various magnetic states, we compute the spin structure factor,

\begin{eqnarray}
 S({\bf q}) = \frac{1}{N}\sum_{ij} e^{-i {\bf q} \cdot ({\bf r}_i - {\bf r}_j)} \langle {\bf S}_i \cdot {\bf S}_j \rangle,
\end{eqnarray}

\noindent
where ${\bf r}_i$ denotes the real-space coordinate of the spin ${\bf S}_i$, $N$ is the number of total spins, and the sum is over all pairs of spins. The spiral states are characterized by a peak in
$S({\bf q})$ at a single value of ${\bf q}$. The corresponding value of ${\bf q}$ is the spiral wave-vector. 
In fact, the nn dot product is another useful quantity in this context. We define a 
general ${\bf Q}$ - average of the dot product as
\begin{eqnarray}
 D_{x}({\bf Q}) & = & \frac{1}{N} \left \langle \left \vert \sum_i e^{i {\bf Q} \cdot {\bf r}_i } {\bf S}_i \cdot {\bf S}_{i+\hat{x}} \right \vert  \right \rangle,
\end{eqnarray}
\noindent
where the verticle bars denote the absolute value, and ${\bf S}_{i+\hat{x}}$ is the nn spin of ${\bf S}_i$ along $x$ direction on the square lattice. One can similarly define $D_{y}({\bf Q})$. 
If the system is in a spiral phase, then $D_{x}({\bf Q})$ at ${\bf Q} = (0,0)$ (denoted as $D_{x}(0)$ here onwards) is finite and serves as an order parameter for the spiral state. 
The $x$ and $y$ components of the spiral wave-vectors can be directly computed as $q_{x/y} = cos^{-1} (D_{x/y}(0))$. Note that a canted state will not be distinguished from a spiral by
using only $D_{x/y}(0)$, therefore it is important to keep track of the spin structure factor $S({\bf q})$. In the EAF phase, $D_{x}(0) = 0$ in the same way as the net magnetization 
is zero in the AFM.

\begin{figure}
\includegraphics[width=.99\columnwidth,angle=0, clip = 'True' ]{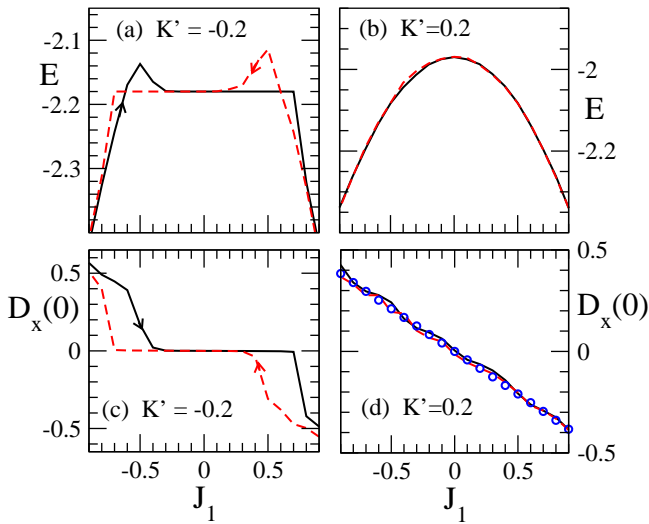}
\caption{(Color online) Energy per spin obtained via Monte-Carlo simulations, plotted as a function of the nn coupling strength $J_1$ for (a) $K' = -0.2$ and (b) $K' = 0.2$. 
Thermally averaged dot product of neighboring spins, $D_x(0)$ (defined in text)
as a function of $J_1$ for (c) $K' = -0.2$ and (d) $K' = 0.2$. The simulations are carried out at $J_{2b} = -1$ and at a fixed temperature $T=0.02$, by increasing (solid lines) and decreasing (dashed lines) the parameter $J_1$. 
The hysteresis in (a) and (c) is an indicator of a first order phase transition, whereas the absence of hysteresis in (b) and (d) suggests a smooth crossover from FM to AFM via the spiral states. Circles in panel (d)
are the corresponding values obtained from variational calculations.
}
\label{fig7}
\end{figure}

We begin by taking a single scan along $J_1$ axis from the phase diagrams presented in Fig. \ref{fig5} at a fixed value of $J_{2b} = -1$. For $K' = -0.2$, the variational phase diagram suggests a transition
from a FM to AFM state via a spiral state and an EAF state. On the other hand, for $K' = 0.2$ the EAF state does not appear. We perform Monte Carlo simulations at constant 
low temperature ($T=0.02$) for the above choice of parameters. We start with a FM state at $J_1 = -2$ and increase the nn coupling constant up to $J_1 = 2$ in steps of $0.02$. The energy per spin and
$D_x(0)$ are tracked as a function of $J_1$. We then begin with an AFM state at $J_1 = 2$ and trace back to $J_1 = -2$. Tracing forward and backward in $J_1$ allows us to probe the hysteresis behavior in the 
system.
The results are presented in Fig. \ref{fig7}. 
Since the order parameter for a spiral state does not evolve continuously to that of an EAF state a hysteresis is present in the computed physical quantities (see Fig. \ref{fig7} (a) and (c)). 
Indeed, according to the variational phase diagram the regime of EAF state is between $-0.6 < J_1 < 0.6$, which is well captured by the Monte Carlo results. The EAF phase is absent in the phase diagram for $K'=0.2$, 
and therefore
the spiral state connects a FM (spiral with wave-vector $(0,0)$) continuously to an AFM (spiral with wave-vector $(\pi,\pi)$), hence no hysteresis behavior is found in Fig. \ref{fig7} (b) and (d). In order to
make these comparisons quantitative, we plot $D_x(0)$ obtained from variational calculations (blue circles in Fig. \ref{fig7}). Monte-Carlo simulations performed directly at low temperature accurately capture the variational
results.

The above checks confirm the validity of ground state phase diagrams obtained via the variational method. They also serve as a benchmark for the efficiency of our Monte-Carlo simulations. 
In addition to the above checks, we also compared the energies obtained from the MC at low temperatures ($T=0.005$) with those 
obtained in the variational calculations, for different choices of parameters belonging to different ground state magnetic phases. These energy comparisons are shown in Table I.  
\noindent
\begin{table}

\begin{tabular}{|l|c|c|}
 \hline

 $ \begin{array}{cc}
   & {Energies} \Rightarrow \\
  {Parameters} \Downarrow &
 \end{array}$                 \ \ \ \  & \ \ \ \ $E_{MC}$  & \ \ \ \ $E_{var}$ \\
 \hline \hline
 $ \begin{array}{c}
    J_{1} = -0.1, 
    J_{2b} = +1.0, 
    K^{'} = -0.2
    \end{array} $     \ \ \ \   &   \ \ \ \   -2.195   & \ \ \ \ -2.200 \\
    
    \hline
    
  $ \begin{array}{c}
    J_{1} = -0.1,
    J_{2b} = -1.0,
    K^{'} = -0.2
    \end{array} $    \ \ \ \    &   \ \ \ \  -2.195   &  \ \ \ \ -2.200 \\
    
    \hline
    
    $ \begin{array}{c}
    J_{1} =  -1.5,
    J_{2b} = -1.0,
    K^{'} = -0.2
    \end{array} $     \ \ \ \  &   \ \ \ \  -3.167   & \ \ \ \ -3.237 \\
    
     \hline
     
     $ \begin{array}{c}
    J_{1} = -0.1,
    J_{2b} = +1.0,
    K^{'} = +0.2
    \end{array} $    \ \ \ \    &  \ \ \ \   -1.995  &  \ \ \ \ -2.000 \\
    
    \hline

\end{tabular}
\caption{Comparison of energy per spin obtained in variational calculations and that obtained in Monte Carlo simulations at low temperature ($T=0.005$), for different choices of parameter sets. The comparison shows that the
variational scheme is able to correctly describe the various ground state phase boundaries within the classical approximation.
}
\label{Table1}
\end{table}
\noindent
The energies compare very well between the Monte Carlo simulations and the variational calculations. We also verify using the spin-structure factor $S({\bf q})$ that the ordered phases obtained in the MC are same as those
shown in the variational phase diagrams for each choice of the parameter sets listed in Table I.

Having established a correspondence between the variational results and the MC simulations for the ground state phase diagrams of $H_2$, we now proceed to describe the finite temperature properties of the model using 
extensive Monte Carlo simulations.
In order to retain the focus, we again refer to the similarities of the present model with those proposed for CuO. Therefore, for the remainder of the paper we focus on a parameter regime which highlights this similarity.

\subsection{Finite temperature results: weakly coupled sub-lattices}

If the hierarchy of the coupling strengths is such that $|J_{2a}| \sim |J_{2b}| >> |J_1| \sim |K'|$, then it is easy to see that the square lattice separates into two sub-lattices which are interpenetrating
each other, but are weakly coupled. In fact, for $K'=0$ these sub-lattices are completely decoupled. While this situation appears only a theoretical possibility in the present model, as the nnn couplings are stronger than 
the nn couplings, it can be realized in 3D oxide structures where the strength and the sign of a superexchange interaction crucially depends on the locations of the bridging oxygens in the crystal structure. 
CuO presents one example, where not only Cu-O-Cu superexchange is important but also the Cu-O-O-Cu super-superexchange contribution is significant \cite{XR}. Such a decoupling into two sub-lattices has been highlighted in
tetragonal CuO, where the intra-sublattice exchange parameters are almost six times larger than the inter-sublattice couplings \cite{Filippetti09}.

\begin{figure}
\includegraphics[width=.99\columnwidth,angle=0, clip = 'True' ]{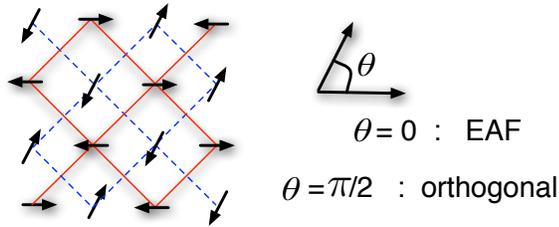}
\caption{(Color online) A schematic view of the decoupled magnetic sub-lattices that arise in the Heisenberg model given by $H_0$ in the limit of small nn coupling strength. Solid and dashed lines indicate the two 
sub-lattices. One can easily check that the ground state energy does not depend on $J_1$ when the order within each sub-lattice is robust, hence any value of the angle $\theta$ between the sub-lattice order parameters 
corresponds to a magnetic ground state. Note that $\theta = 0$ and $\theta = \pi/2$ correspond to the collinear EAF state and the orthogonal spiral state, respectively.
}
\label{fig8}
\end{figure}

We begin by analyzing the effects of biquadratic coupling on the ground state in a simple variational picture. Lets assume that the vector order parameters of the decoupled sub-lattices make an angle $\theta$ with each 
other. This is schematically shown in Fig \ref{fig8} where solid (red) and dashed (blue) lines highlight the two sub-lattices. Within each sub-lattice the spins are arranged ferromagnetically along one diagonal and
antiferromagnetically along the other. The relative orientation of the spins on two sub-lattices is parameterized by angle $\theta$.
The ground state energy of the system in the absence of bi-quadratic couplings is independent of $\theta$, and hence
there is a degeneracy of magnetic states in the system. The additional energy contribution due to the biquadratic coupling $K'$ can be written as $E(\theta) = K' \cos^2 \theta $. Minimizing the energy by demanding
$dE/d\theta = 0$ and $d^2E/d\theta^2 > 0$ , leads to two possible solutions: $\theta_{min} = 0$ for $K' > 0$ and  $\theta_{min} = \pi/2$ for $K' < 0$. 
Therefore the introduction of a biquadratic coupling term lifts the degeneracy and stabilizes two specific phases, the EAF state or the orthogonal spiral state, depending on the sign of $K'$. 
These two phases are analogous to the AF$_1$ and AF$_2$ magnetic states of CuO. 
One of the interesting features in CuO is the existence of a non-collinear AF$_2$ magnetic state at high temperatures. This magnetic state supports a finite spin current due to its spiral nature and therefore
is responsible for the high-temperature ferroelectric behavior of CuO. This is unusual as in most frustrated magnets, it is the low-temperature phase that supports a non-collinear magnetic phase and hence the
ferroelectricity is observed only at low temperatures. Given the similarity of the present model with that of CuO, it is natural to ask if a spiral magnetic order exists at high temperatures within the present 
model.

\begin{figure}
\includegraphics[width=.9\columnwidth,angle=0, clip = 'True' ]{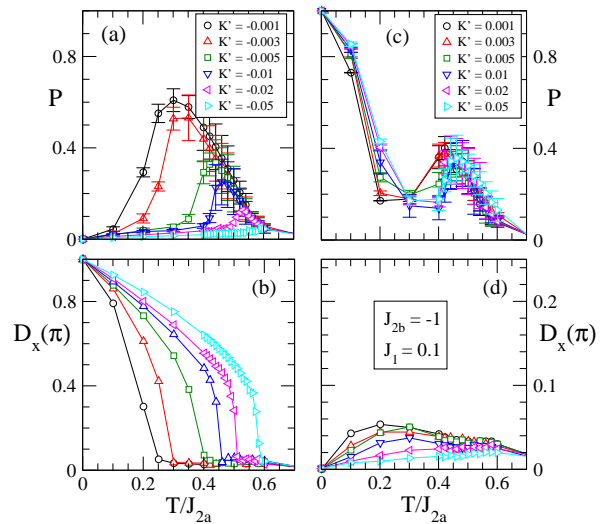}
\caption{(Color online) Temperature dependence of the spin current $P$ for different values of $K'$ in the range, (a) $K' < 0$ and (c) $K' > 0$. Temperature dependence of the order parameter $D_x(\pi)$ for the 
collinear magnetic state EAF
for various values of $K'$ in regimes, (b) $K' < 0$ and (d) $K' > 0$. Results are obtained on $120 \times 120$ lattice, with $J_1 = 0.1$ and $J_{2b} = -1$. Note the difference in scale on y-axis in panel (d).
}
\label{fig9}
\end{figure}

In order to answer this, we compute the spin current as a measure of the spiral nature of non-collinear magnetism. Note that a canted phase is also non-collinear but this will not lead to a finite spin current.
Following Katsura {\it et al.} we define spin current as \cite{Katsura}

\begin{eqnarray}
P & = & \frac{1}{N}\left \langle \left \vert \sum_{i} {\bf P}_i \right \vert \right \rangle  \nonumber \\
 {\bf P}_i & = & {\bf \hat {x}} \times ({\bf S}_i \times {\bf S}_{i+\hat{x}}), 
\end{eqnarray}

\noindent
where the vertical bars denote the magnitude of the vector and the angular brackets denote thermal average.
$\vert {\bf P}_i \vert$ can be regarded as a local measure of spin current on a given bond along x-direction originating at lattice site $i$. 
Temperature dependence of the spin current $P$ is shown in Fig. \ref{fig9}(a) for negative values of $K'$. 
The Monte Carlo results are consistent with the variational analysis and $P$ indeed vanishes at low temperatures.
The high temperature behavior is intriguing: there is a window in temperature where $P$ becomes finite, 
indicating the existence of a non-collinear magnetic state at finite temperatures.
Interestingly, the temperature range of stability of the non-collinear magnetic state increases upon decreasing the magnitude of $K'$. This suggests that the non-collinear phase may be entropy driven, as a 
large negative $K'$ favors the collinear state and hence competes with the non-collinear state at high temperatures, narrowing the window of stability of the later state.
We also notice that the spin current shows large fluctuations, which is the origin of relatively large error bars in Fig. \ref{fig9}(a),(c).

Recalling the definition Eq. (13), we note that $D_x({\bf Q})$ at ${\bf Q} = (\pi,\pi)$ (denoted as $D_x(\pi)$) is an order parameter for the EAF phase.
${\bf S}_i \cdot {\bf S}_{i+\hat{x}}$ represents the local degree of collinearity between neighboring spins along $x$-direction. Note that in the EAF phase, the dot-product between
neighboring spins have a perfect staggered order. This motivates the use of $D_x(\pi)$ as an order parameter for the EAF phase, much like the staggered magnetization is an order parameter for the AFM state.
The temperature-dependence of
$D_x(\pi)$ is plotted in Fig. \ref{fig9} (b) for $K' < 0$. The ground state is collinear as the order parameter rises smoothly to its maximal value upon reducing temperature
in our simulations. The temperature at which $D_x(\pi)$ shows an abrupt rise, correlates well with the temperature at which the spin current begins to fall.
Therefore, the temperature dependence of $P$ and $D_x(\pi)$ can be used to assign characteristic temperatures at which the collinear and the spiral magnetic orders set in.
Next, we study the behavior of the model for positive values of $K'$. In agreement with the variational results, the ground state supports a non-collinear order (see Fig. \ref{fig9} (c)). $P=1$ indeed
corresponds to the orthogonal magnetic phase where neighboring spins point at an angle of 90$^{\circ}$ to each other. 
Therefore the transition between a collinear and non-collinear magnetic orders as a function of $K'$ is well captured in our simulations. However, the finite-temperature behavior of the
model is quite different for positive $K'$. The spin current shows a non-monotonic dependence on temperature, which is similar for different values of $K'$. Upon increasing temperature
$P$ decreases smoothly from its maximum value and 
then shows a weak rise before decreasing again. The collinear order parameter does not show any considerable rise in the intermediate temperature range.
For $N=120^2$, we find that $D_x(\pi) \sim 0.02$ in the high-temperature PM state. The maximum value of $D_x(\pi)$ in Fig. \ref{fig9}(d) is only about two times larger. This is in sharp contrast with the
behavior of spiral order parameter $P$, which in the negative $K'$ regime shows a value about 30 times larger than that in PM state (see Fig. \ref{fig9}(a)).
Clearly, the collinear state is only energetically stable and exists at finite temperatures only when $K'<0$. Therefore, our results suggest that the non-collinear phase at high temperatures is stabilized
by entropic effects. This is in agreement with previous material-specific models for CuO \cite{CuO, He}. However, our results show that these results are a general feature of a model where
the standard Heisenberg term leads to a degeneracy of magnetic states, and the ground state selection relies on weaker higher-order couplings. One clear advantage of the bilinear-biquadratic model is that the 
competition is only between EAF and orthogonal spiral states. This is unlike the models invoking DM coupling and magnetocrystalline anisotropy where it is possible to stabilize
a spiral phase with arbitrary ${\bf q}$ depending on the relative strengths of the DM coupling and anisotropy \cite{CuO}. 

\begin{figure}
\includegraphics[width=.99\columnwidth,angle=0, clip = 'True' ]{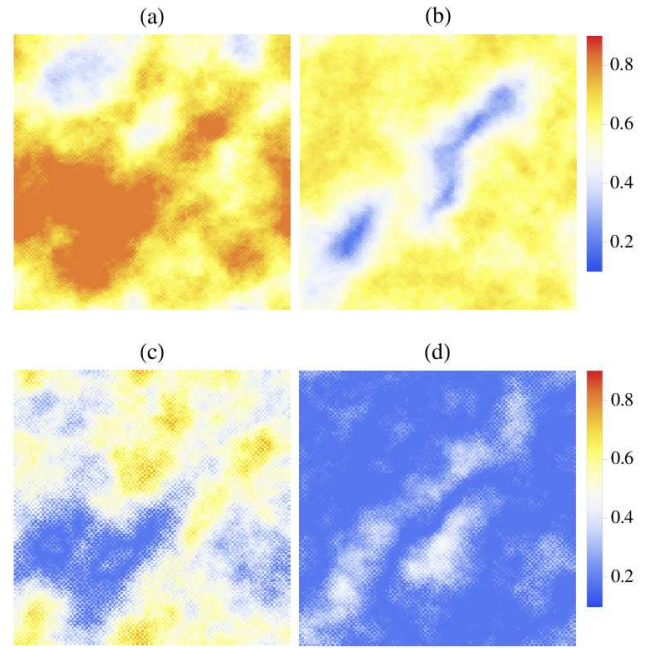}
\caption{(Color online) Real-space patterns of local measures for non-collinear and collinear orders, $P_i$ (upper row) and ${\bf S}_i \cdot {\bf S}_{i+\hat{x}}$ (lower row), 
for $K' = -0.005$ ((a) and (c)) and $K' = 0.005$ ((b) and (d)) at intermediate 
values of temperature $T=0.4$. These are single configurations that contribute to the averages plotted in Fig. \ref{fig9}.
}
\label{fig10}
\end{figure}

One of the advantages of the real-space approach is that we can investigate the spatial character of the finite-temperature non-collinear magnetism.
For this purpose, we plot the real-space snap-shots of the local collinear and non-collinear order parameters, ${\bf S}_i \cdot {\bf S}_{i+\hat{x}}$ and $P_i$.
For $K' < 0$, the ground state is the up-up-down-down type antiferromaget which corresponds to $\theta = 0$ in Fig. \ref{fig8} (denoted as EAF). 
The real space pattern at finite temperature ($T=0.4$) show regions with strong non-collinear spin order (see Fig. \ref{fig10}(a)).
For $K' > 0$, the ground state supports a perfect orthogonal spiral order, where neighboring spins make an angle of $\pi/2$ with each other. At finite temperatures, regions with weak spiral order are present as inferred from
the real-space distribution of $P_i$ (see Fig. \ref{fig10}(b)). Fig. \ref{fig10}(c),(d) show the corresponding snap-shots of the collinear order, ${\bf S}_i \cdot {\bf S}_{i+\hat{x}}$. 
Collinear order is absent at finite temperatures. While
we only show $P_i$ and ${\bf S}_i \cdot {\bf S}_{i+\hat{x}}$ for a single configurations here, we have verified that these are typical configurations for $T=0.4$. 
These real-space patterns also highlight the asymmetry w.r.t. $K'$ that we discussed earlier.
While at low 
temperatures the sign of $K'$ decides the nature of the ordered magnetic state, at finite 
temperatures it is the non-collinear state that dominates irrespective of the sign of $K'$.
At finite temperatures there are many different configurations of spins that are involved in the final thermal average. The spin current shows strong fluctuations at intermediate temperatures as we noticed 
in Fig. \ref{fig9}(a),(c). However, the energy 
as a function of temperature (not shown) behaves smoothly with error bars less than the typical symbol size in our figures. Strong fluctuations in spin current with weak fluctuations in total energy indicate an unusual 
energy landscape in configuration space where many 
different configurations are present within a narrow energy range, and the system easily accesses these configurations. This explains why the spiral order wins over collinear order at finite temperatures: the availability 
of many different magnetic configurations with finite spin current increases the entropy of the system. Therefore, the system prefers to visit such configurations more often leading to a finite spin current. This
highlights a key role of entropy in stabilizing the non-collinear magnetic order at finite temperatures. Moreover, our results suggest that the finite temperature spiral phase is inhomogeneous. 
This compares favorably with recent 
experimental investigations, where the presence of an inhomogeneous ferroelectric phase in CuO has been reported \cite{nanoscale_CuO}.

\begin{figure}
\includegraphics[width=.85\columnwidth,angle=0, clip = 'True' ]{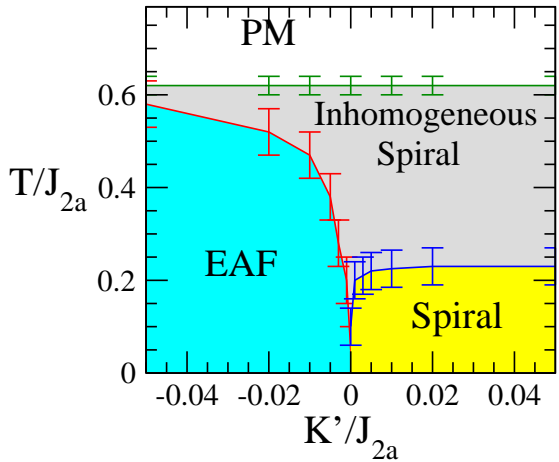}
\caption{(Color online) Temperature versus $K'$ phase diagram for $J_{2b} = -1$ and $J_1 = 0.1$. The inhomogeneous spiral phase consists of islands of spiral phase coexisting with paramagnetic regions.
}
\label{fig11}
\end{figure}

We summarize the results discussed in this section in a finite-temperature phase diagram. Clearly, in the thermodynamic limit there is no long-range order at non-zero temperatures in two-dimensions for a 
classical continuous-spin model \cite{Mermin-Wagner, Gelfert}. 
Nevertheless, numerical simulations on finite lattices can be used to identify characteristic temperatures at which the correlation length for a
particular magnetic order exceeds the system size. If the model is extended to three dimensions by allowing a weak interplanar coupling between the 2D layers, the characteristic temperature
scales discussed here will be relevant to the genuine phase-transition temperatures. We use the temperature-dependence of $D_x(\pi)$ and $P$ (as shown in Fig. \ref{fig9}) to identify such characteristic 
temperature scales for collinear EAF and spiral orderings, respectively. These are brought
together as a phase diagram in Fig. \ref{fig11}. The ordering between the PM and the inhomogeneous spiral phase is independent of $K'$. This corresponds to the onset of ordering of the two sub-lattices (see Fig. \ref{fig8}), 
which is decided by the 
strengths of the couplings $J_{2a}$ and $J_{2b}$. The inhomogeneous spiral phase corresponds to a state where PM regions coexist together with regions supporting large values of spin current. 
Upon reducing temperature further, the
system orders into a collinear or non-collinear state depending on the sign of $K'$. 
The similarity of our results with those obtained on material-specific models for CuO is apparent \cite{CuO, He}. The experimental situation of CuO is captured well in our model if we assume a
small negative value of the biquadratic coupling $K'$, where
a window of stability of the non-collinear state exists between the collinear EAF and the fully disordered PM state. 
Our results suggest that this intermediate spiral phase contains inhomogeneities with patches of spiral patterns coexisting
with magnetically disordered regions.

\section{Conclusions}

We use a combination of variational calculations and Monte Carlo simulations to study
a classical spin model on a 2D square lattice with nearest neighbor and next nearest neighbor Heisenberg exchange interactions and nearest neighbor biquadratic interactions. Motivated
by the existence of a non-collinear magnetic state at high temperatures in CuO \cite{CuO-Kimura}, we study this model in a specific parameter regime where the nnn couplings are inequivalent and are much larger than the
nn exchange interactions. In this limit, the system is decoupled into two sub-lattices and lead to a situation where even a weak biquadratic coupling is vital in selecting the ground state magnetic order.
We also compare the ground states of the bilinear-biquadratic model with those of a model containing a four-spin ring exchange coupling. 
It is the ring-exchange coupling that is
physically relevant for a spin-$1/2$ system such as CuO. However, we demonstrate that the classical versions of the quantum models for spin-$1/2$ and spin-$1$ have 
strong similarities in the parameter regime considered here. The finite-temperature study of the bilinear-biquadratic model using Monte Carlo 
uncovers an inhomogeneous non-collinear 
magnetic state. The $T-K'$ phase diagram of the model captures the main features of the phase diagrams reported in material-specific models for CuO, which considered the role of DM coupling and weak anisotropy.
The present work helps in understanding the finite-temperature magnetism in CuO within a pure spin model in a general setting without taking into account the details of the
crystal structure.
We have shown that if the stronger interactions are such that they compete and lead to a degeneracy of magnetic
ground states, then the role of the much weaker higher-order interactions cannot be ignored. These interactions are as important as other weak effects like the magnetocrystalline anisotropies and the spin-orbit 
induced DM interactions. Using the real-space analysis we also find 
that the high-temperature spiral phase is inhomogeneous with islands of spiral phase coexisting with paramagnetic regions. The common feature with the previous model studies is that the entropic effects 
are crucial in stabilizing the non-collinear state at high temperatures \cite{CuO,He}.
While there have already been some experimental reports that point to an inhomogeneous ferroelectric state in CuO \cite{nanoscale_CuO}, it would be very interesting to probe further the
spatial nature of the high-temperature non-collinear phase in CuO.

\subsection{ACKNOWLEDGMENTS}
We acknowledge the use of High Performance Computing Facility at IISER Mohali. 
K.P. acknowledges support via CSIR/UGC fellowship.
S.K. acknowledges insightful discussions with Zohar Nussinov and Prabuddha Chakraborty, and financial support from DST, India.

\end{document}